\documentclass[aps,twocolumn]{revtex4-1}
\usepackage{graphicx,epsfig}
\usepackage{amsmath}
\usepackage[T1]{fontenc}
\usepackage{hyperref}
\hypersetup{
     colorlinks   = true,
     citecolor    = magenta,
     linkcolor    = blue,
}

\begin{document}

\title{Spin-1 bosons in an
  external magnetic field and a three body interaction potential}
\author{Sk Noor Nabi}
\author{Saurabh Basu} \email{saurabh@iitg.ernet.in} \affiliation{Department of
    Physics, Indian Institute of Technology Guwahati, Guwahati, Assam
    781039, India} \date{\today}

\begin{abstract}
We perform a thorough study of the effect of an external magnetic field
on a spin-1 ultracold Bose gas via mean field approach
corresponding to the both signs of the spin dependent interaction. In contrast to some of the earlier studies, the magnetic field in our work is included through both the hopping frequencies (via Peierls coupling) and the zeeman interaction, thereby facilitating an exploration for competition between the two. The phase
diagrams in the antiferromagnetic case shows that the Mott insulating
(MI) phase with even particle occupancies is stable at low magnetic
fields. At higher magnetic fields, due to a competition between the
hopping and the zeeman interaction terms, the latter tries to
destabilize the MI phase by suppressing the formation of singlet
pairs, while the former tends to stabilize the MI phase. In the
ferromagnetic case, the MI lobes become more stable with increasing
flux strengths. Further inclusion of a three body interaction
potential in order to ascertain its role on the phase diagram, we found that
in absence of the magnetic field, the MI lobes
become more stable compared to the superfluid (SF) phase and the
location of the transition point for the MI-SF phase increases with
increasing the three body interaction strength. A strong coupling
perturbative calculation has also been done to provide a comparison with our
mean field phase diagrams. Lastly, with inclusion of the external field, the
insulating phases are found to be further stabilized by the three body
interaction potential.
\end{abstract}


\maketitle

\section{Introduction}
\label{intro}
The experimental realization of the transition from a SF to MI in a
system of neutral alkali atoms trapped in optical lattices has marked
an important milestone towards exploring the many body phenomena in
systems which demonstrate quantum phase transition \cite{Greiner}. Usage of two or more counter propagating laser beams, which form the
optical lattice potential, allows one to have a precise control of the
lattice parameters and the interaction potential between the
constituent particles experimentally. These technological advancements have made the scientists capable
to navigate through a myriad of interesting physical phenomena that
are otherwise inaccessible in crystalline solids. The cooling of atoms
involves sophisticated trapping techniques that are either magnetic or
optical in nature.  \\ \indent In magnetic trapping, ultracold atoms
are forced to have their internal atomic states frozen and hence
behaves like a spin-0 or a scalar Bose gas significantly missing the
rich phase properties. While in optical trapping, the interaction
between the electric field of the laser beams and the dipole force of
the neutral atoms favors in retaining the hyperfine spin states. Thus
the system can be treated as spinor Bose gas which shows a plethora of
interesting phase properties compared to the scalar Bose 
gas \cite{PhysRevLett.80.2027,PhysRevLett.81.742,ohmi,PhysRevA.68.043612,Uesugi,PhysRevLett.81.5257,PhysRevLett.88.163001,PhysRevLett.94.110403}.
\\ \indent The dynamics of the ultracold atoms loaded in optical lattices
were first theoretically analyzed by the well known Bose Hubbard Model
(BHM) in a seminal paper by Jaksch $et.$ $al.$
\cite{PhysRevLett.81.3108} where the SF-MI phase transition is found
to be completely governed by the competition between the hopping and
two body interaction strengths. Following this, different variants of
BHM with nearest neighbor extended interactions
\cite{PhysRevLett.94.207202,PhysRevA.80.043614,PhysRevA.83.051606,PhysRevB.86.054520,apurba,Apurba1},
three and higher body interaction strengths
\cite{PhysRevLett.101.150405,PhysRevA.78.043603,
  PhysRevA.88.063625,PhysRevA.90.041602,PhysRevA.92.043615}, disorder
\cite{PhysRevA.81.063643,PhysRevA.91.043632,PhysRevA.91.031601} and
multicomponent mixture of Bose gas
\cite{PhysRevA.67.013606,PhysRevA.92.053610}, superlattice potential
\cite{PhysRevA.81.053608,PhysRevA.85.051604} etc have been studied. These studies
nowadays contributes significantly for the exploration of the quantum
gases. Needless to say disorder (or other inhomogenities) play an
important role in shaping the physics of such systems. \\ \indent The
general properties of spinor Bose condensates was first studied by Ho
\cite{PhysRevLett.81.742} and at the same time by Ohmi $et.$ $al.$ \cite{ohmi}
where the system is characterized by a vector order parameter. The
vector property of the condensate hence shows significant modification
of the ground state structures and yields new topological excitations
as compared to its scalar component. Later several studies on spinor
Bose gas include the evolution of spin and singlet order parameters
\cite{PhysRevA.91.043620}, spin orbit coupling
\cite{PhysRevA.86.043602,PhysRevA.93.013629,PhysRevA.93.023615,PhysRevB.93.081101},
effects of disorder \cite{PhysRevA.83.013605,Noor} etc now under the
lens from theoretical as well as experimental perspectives.  \\\indent
Recently due to the hyperfine spin states of spinor Bose gas, the
creation of synthetic gauge fields and the observation of quantum Hall
effect are of great importance where these zeeman sub levels act as a
synthetic dimensions along the short axis against the optical lattice
sites along the long axis \cite{PhysRevLett.112.043001,Stuhl1514}.
\\ \indent Besides that there are large number of literature review on
spinor Bose gas under external magnetic fields which include the study
of phase diagrams \cite{PhysRevA.68.043612,Uesugi}, statistical
physics of spin dynamics at finite temperature
\cite{PhysRevA.72.023617,PhysRevA.90.043609}, spatial and spin
structures of ground state \cite{PhysRevA.82.053630}, phase separation
\cite{PhysRevA.80.023602}, exact eigenstates
\cite{PhysRevLett.84.1066} etc. Motivated by the studies carried out
in Refs.\cite{PhysRevA.68.043612,Uesugi}, where they consider the
effect of external magnetic field only through the zeeman interaction
term and show that the MI phase destabilizes with increasing field
strength, here we consider the effect of the magnetic fields on
both the hopping (via Peirls coupling) and the zeeman interaction terms for a chosen (Landau)
gauge and see their competition on the MI-SF phase transition.
\\\indent Apart from considering the effects of magnetic field, we
feel it should be interesting to see the effects of three body
interaction on spinor Bose gas, a topic which has not got enough
attention. Unlike a scalar Bose gas,
for a spinor Bose gas, the three body interaction strengths consists
of two parts namely as, the three body spin independent and another
one which is the spin dependent interaction terms as derived in Ref.\cite{PhysRevA.88.023602}.  \\ \indent In section II, we outline our
theoretical model for a spinor Bose gas in presence of magnetic field
and study it via the familiar mean field technique. In section III, we
discuss the phase diagrams of the system in presence of the magnetic
field and hence include a three body interaction potential to study an
interplay between them. Finally we draw our conclusions in section IV.
\section{Model}
The Hamiltonian for spin -1 ultracold atoms in presence of a magnetic
field, $B$ pointing in the $z$ direction and enters through a vector
potential ${\bf{A}}$ chosen in the Landau gauge as,
${\bf{A}}=Bx\hat{y}$, is written as
\cite{PhysRevLett.81.742,ohmi,PhysRevA.68.043612,Uesugi,PhysRevB.60.2357},
\begin{eqnarray}
H&=&-\sum\limits_{<ij>}\sum\limits_{\sigma}(t_{i,j}a^{\dagger}_{i\sigma}a_{j\sigma}+h.c)-\mu
\sum\limits_{i} n_{i}+\frac{U_{0}}{2}\sum\limits_{i}n_{i}\nonumber \\ &&(n_{i}-1)+\frac{U_{2}}{2}\sum\limits_{i}({\bf{S}}^{2}_{i}-2n_{i})+g\mu_{B}B\sum\limits_{i} S_{iz}
\label{bhm1}
\end{eqnarray}
Here $t_{ij}$ is the hopping matrix elements from site $i$ to nearest
neighbour site $j$ and is related to the magnetic vector potential $\bf{A}$
via Peierls coupling as
$t_{ij}=te^{-i\frac{2\pi}{\phi_{0}}\int\limits_{i}^{j}\bf{A}.\bf{dl}}$
where $\phi_{0}$ is the magnetic flux quantum. For a particular
choice of the gauge field, the integral assumes the value as $\phi=B
l^{2}_{0}/\phi_{0}$, $l_{0}$ being the lattice
spacing. $a^{\dagger}_{i\sigma}(a_{i\sigma})$ is the boson creation
(annihilation) operator at a site $i$ and the particle number operator
is $n_{i}=\sum\nolimits_{\sigma}n_{i\sigma}$,
$n_{i\sigma}=a_{i\sigma}^{\dag}a_{i\sigma}$. $U_{0}$ is the spin
independent and $U_{2}$ is the spin dependent on-site interactions
which are related to the scattering lengths $a_{0,2}$ by $U_{0}=(4\pi
\hbar^{2}/M)((a_{0}+2a_{2})/3)$ and $U_{2}=
(4\pi\hbar^{2}/M)((a_{2}-a_{0})/3)$ corresponding to $S$=0 and $S$=2
channels respectively \cite{PhysRevLett.81.742,ohmi}. If the spin dependent
interaction is $U_{2}/U_{0}>0$, then it is known as antiferromagnetic
(AF) interaction and for $U_{2}/U_{0}\le 0$, it is known as ferromagnetic
interaction. The total spin at a site $i$ is given by,
${\bf{S}}_{i}=a^{\dagger}_{i\sigma}{\bf{F}}_{\sigma\sigma'}a_{i\sigma'}$
where ${\bf{F}}_{\sigma\sigma'}$ are the components of spin-1 matrices
and $\sigma=$+1, 0, -1. $\mu$ is the chemical potential, $g$ is Lande
g factor and $\mu_{B}$ is the Bohr magneton and
$S_{iz}=a^{\dagger}_{i+}a_{i+}-a^{\dagger}_{i-}a_{i-}=n_{i+}-n_{i-}$
is the $z$ component of ${\bf{S}}_{i}$. Here we consider a two
dimensional square lattice where every lattice site $i$ can be
expressed by two indices as $i=[l,m]$, $l$ corresponds to lattice site
along $x$ direction and $m$ along $y$ direction of the lattice.
\\ \indent To decouple the hopping term, we use the mean field
approximation where the hopping term can be written as
\cite{PhysRevB.75.045133,PhysRevB.77.014503},
\begin{equation}
\begin{aligned}
a^{\dagger}_{(l,m)\sigma}a_{(l+1,m+1)\sigma}&\simeq \langle
a^{\dagger}_{(l,m)\sigma} \rangle a_{(l+1,m+1)\sigma}\\
&+a^{\dagger}_{(l,m)\sigma}\langle a_{(l+1,m+1)\sigma}\rangle
\end{aligned}
\end{equation}
where $\langle\hspace{2mm} \rangle$ denotes the equilibrium value of
an operator. Defining the superfluid order parameter at a site $(l,m)$
as, $\psi_{(l,m)\sigma}=\langle a_{(l,m)\sigma}\rangle$, the total SF
order parameter is given by
$\psi_{l,m}=\sqrt{\psi^{2}_{(l,m)\sigma}}$. Substituting this in
Eq.(\ref{bhm1}), the BHM can be written as a sum of single site
Hamiltonians as, $H=\sum\nolimits_{l,m}H^{MF}_{l,m}$ where,
\begin{eqnarray}
H^{MF}_{l,m}&=&-\sum\limits_{\sigma}[t_{l+1,m}\psi^{*}_{l+1,m}a_{l,m}+t_{l-1,m}\psi^{*}_{l-1,m}a_{l,m}\nonumber
  \\&+&t_{l,m+1}\psi^{*}_{l,m+1}a_{l,m}+t_{l,m-1}\psi^{*}_{l,m-1}a_{l,m}+h.c]\nonumber\\ &+&\frac{U_{0}}{2}
n_{l,m}(n_{l,m}-1)+\frac{U_{2}}{2}({\bf{S}}^{2}_{l,m}-2n_{l,m})\nonumber
\\ &-&\mu n_{l,m}+\eta [n_{(l,m)+}-n_{(l,m)-}]
\label{mf}
\end{eqnarray}
where $\eta=g\mu_{B} B$. Using the Bloch periodic boundary condition
and calculating the hopping matrix element between site $(l,m)$ to a
nearest neighbour $(l\pm 1, m\pm 1)$, we can write,
\begin{equation}
t_{l\pm 1,m\pm 1}\psi_{l\pm 1,m\pm 1}=
\begin{cases}
t\psi_{l\pm 1,m}; l,m=l\pm 1,m\\
te^{\mp i 2\pi l \phi}\psi_{l,m}; l,m=n,m\pm 1
\end{cases}
\end{equation}
Now to compute the ground state energy of the system, we first
evaluate the matrix elements of the mean field Hamiltonian,
$H^{MF}_{l,m}$ in the occupation number basis,
$|n_{(l,m)\sigma}\rangle$ as,
\begin{eqnarray}
\langle\hat{n}_{l,m+},\hat{n}_{l,m0},\hat{n}_{l,m-}|H^{MF}_{l,m}|\hat{n}'_{l,m+},\hat{n}'_{l,m0},\hat{n}'_{l,m-}\rangle&=&\nonumber\\
h^{d}_{l,m}+h^{od}_{l,m}
\label{mf1}
\end{eqnarray}
where the $h^{od}_{l,m}$ correspond to the matrix elements coming
from the off diagonal terms as, 
\begin{equation}
h^{od}_{l,m}=-t\sqrt{n_{l,m}}[\psi_{l+1,m}+\psi_{l-1,m}+(e^{-i2\pi \phi
    l}+ e^{i2\pi\phi l})\psi_{l,m}]
\end{equation}
and the diagonal part, $h^{d}_{l,m}$ is calculated as
\begin{eqnarray}
h^{d}_{l,m}&=&\frac{U_{0}}{2}
n_{l,m}(n_{l,m}-1)+\frac{U_{2}}{2}({\bf{S}}^{2}_{l,m}-2n_{l,m})\nonumber
\\ &-&\mu n_{l,m}+\eta [n_{(l,m)+}-n_{(l,m)-}]
\end{eqnarray}
After diagonalizing Eq.(\ref{mf1}) with $n=7$ for which
$\langle$$\hat{n}_{l,m+},\hat{n}_{l,m0},\hat{n}_{l,m-}|H^{MF}_{l,m}|\hat{n}'_{l,m+},\hat{n}'_{l,m0},\hat{n}'_{l,m-}$$\rangle$
is a $120\times120$ matrix, we obtain the ground state energy,
$E_{g}(\psi_{l,m\sigma})$ and the eigenfunctions,
$\Psi_{g}(\psi_{l,m\sigma})$ starting with some some guess value for
$\psi_{l,m\sigma}$. Now from the updated wave function
$\Psi_{g}(\psi_{l,m\sigma})$, we compute the equilibrium SF order
parameter and local densities self consistently using,
\begin{eqnarray} 
\psi_{l,m\sigma}&=&\langle\Psi_{g}(\psi_{l,m\sigma})|a_{l,m\sigma}|\Psi_{g}(\psi_{l,m\sigma})\rangle
\\ \rho_{l,m}&=&\langle\Psi_{g}(\psi_{l,m\sigma})|n_{l,m\sigma}|\Psi_{g}(\psi_{l,m\sigma})\rangle
\end{eqnarray}
\indent It is relevant to mention that in absence of the magnetic
field, the Hamiltonian is site independent and the SF order
parameters are uniform over all the lattice sites, while in presence
of magnetic field, they are site dependent and show a direct SF-MI
phase transition caused by a competition among the hopping and
interaction strengths. In the strong interaction limit, the system is
in the MI phase which is basically a random phase with a vanishing SF
order parameter and fixed number of bosons per lattice site. While at
lower interaction strengths, the system switches over to the the conducting
phase, that is the SF phase with finite SF order parameter and non
integer occupation densities which can be perceived as an ordered phase.
\section{Results}
\subsection{Magnetic field}
\begin{figure}[!ht]
  \centerline{ \hfill
    \psfig{file=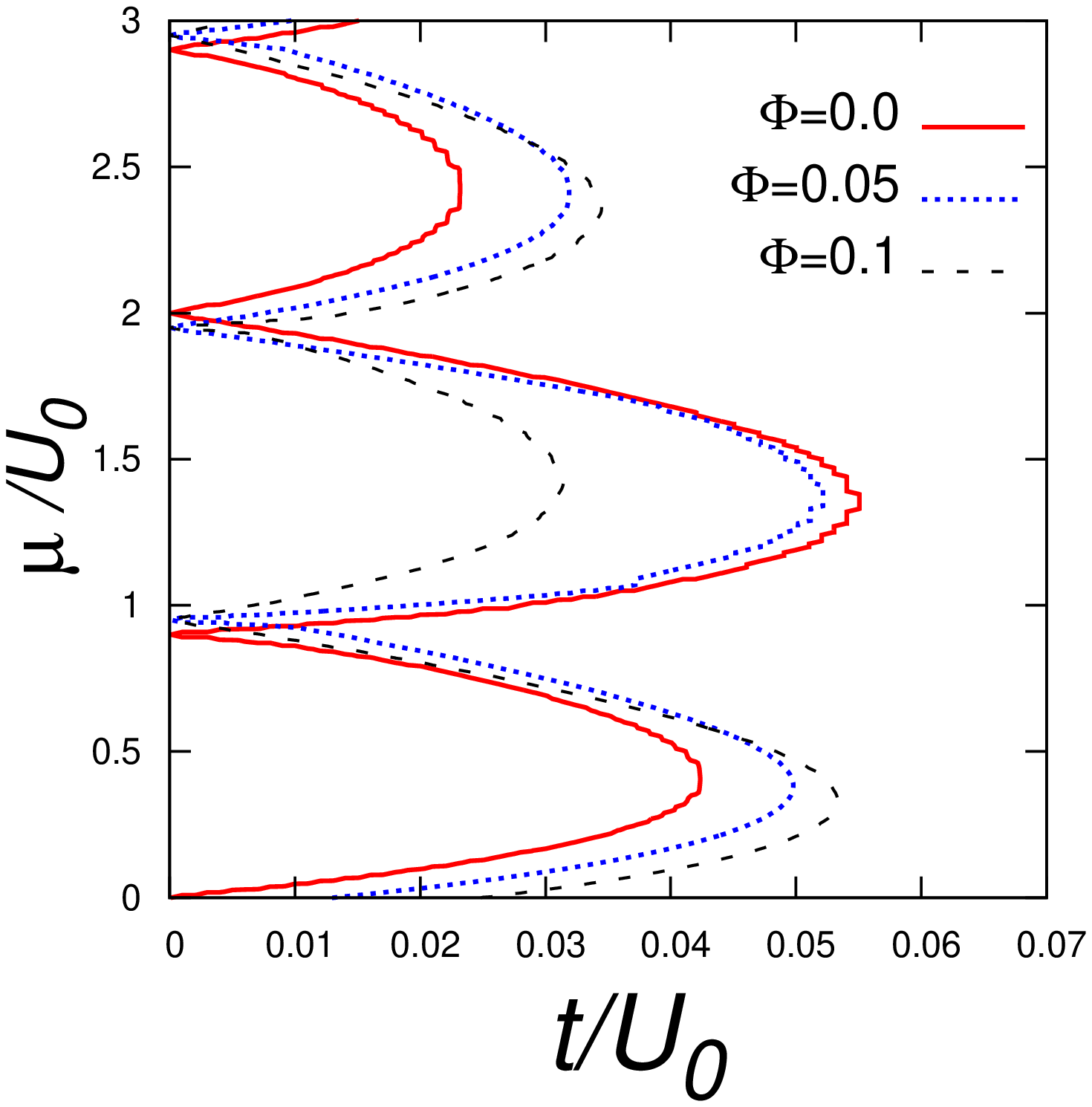,width=0.25\textwidth}
    \hfill \hfill \psfig
           {file=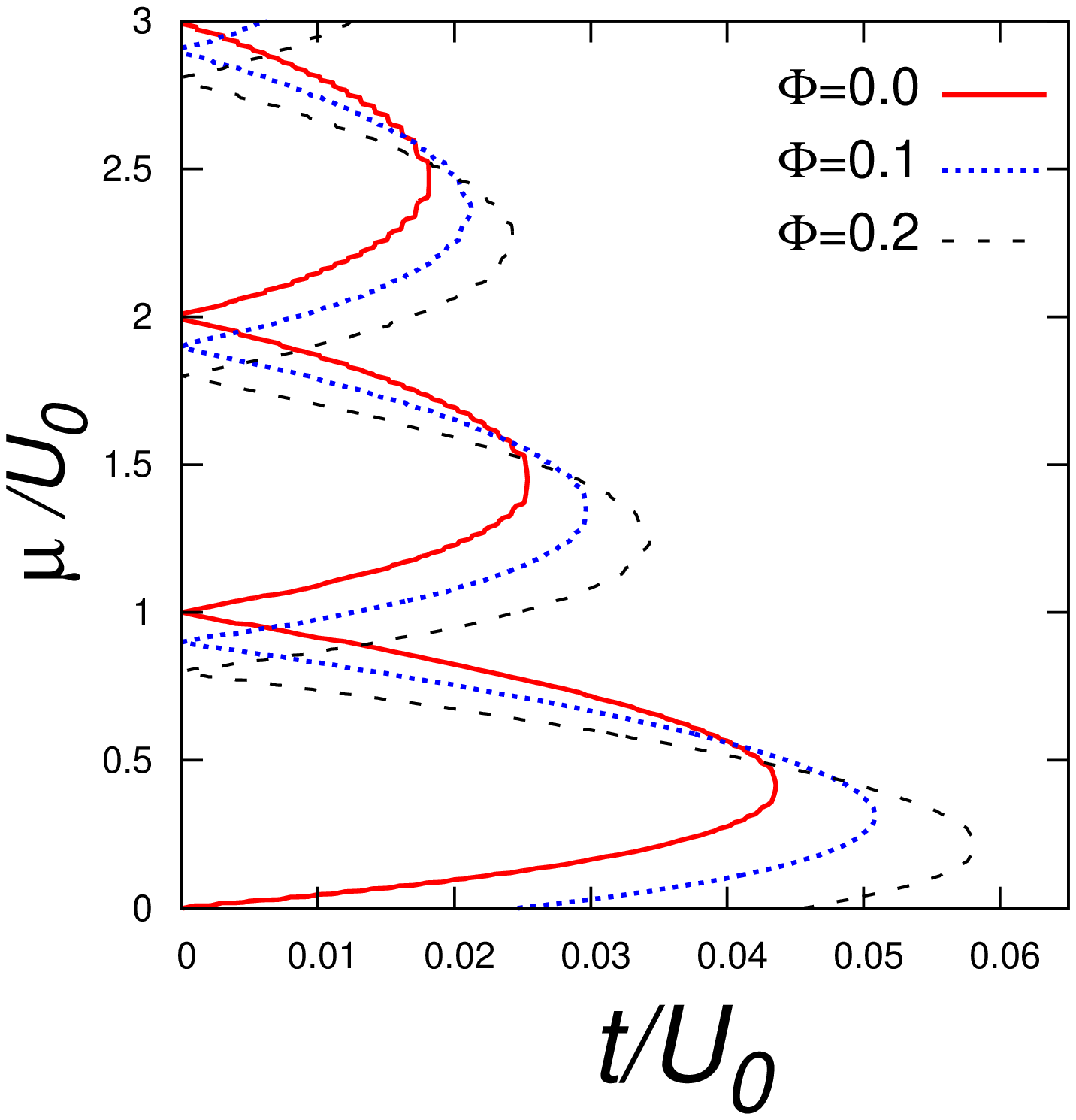,width=0.25\textwidth}
           \hfill} \centerline{\hfill $(a)$ \hfill\hfill $(b)$ \hfill}
  \caption{(Color online) Phase diagrams for different values of
    $\phi$ in antiferromagnetic case with $U_{2}/U_{0}=0.05$ in (a)
    and for ferromagnetic case with $U_{2}/U_{0}=0.0$ in (b). In AF
    case, at higher magnetic flux, the even MI lobes becomes unstable
    while odd MI lobes occupy more regime compared to the SF phase. }
\label{polarferromag}
\end{figure}

Here we consider bosons in a magnetic field by choosing the magnetic
flux to be expressible in the form of a rational fraction, that is,
$\phi=p/q$ where $p,q$ are integers
\cite{PhysRevB.14.2239,PhysRevB.60.2357}. In the chosen Landau gauge,
from Eq.(\ref{mf1}), the system is translationally invariant along the $y$
direction and quasi periodic (as explained below) in the $x$
direction. For a two dimensional square lattice with lattice site
indices as $(l,m)$, the system maintains its periodicity along $x$
direction with period $l=q$. This implies that we shall have to
diagonalize the mean field Hamiltonian over a one dimensional
chain of length $q$, that is, a $1\times q$ magnetic supercell with
periodic boundary conditions in the $x$ direction in order to obtain the ground
state energy. Due to invariance along the $y$ axis, the SF order
parameter, $\psi_{l,m}$ and occupation density, $\rho_{l,m}$ are
independent of $m$ that is $\psi_{l,m}=\psi_{l}$ and
$\rho_{l,m}=\rho_{l}$. Further owing to the periodicity in the
$x$-direction, we also have $\psi_{l}=\psi_{l+q}$ and
$\rho_{l}=\rho_{l+q}$. We shall obtain the phase diagram based on the
site averaged SF order parameter, $\bar\psi=\sum\nolimits_{q}\psi(q)/q$
and local density, $\bar\rho=\sum\nolimits_{q}\rho(q)/q$ for different
values of the magnetic flux, $\phi$ \cite{PhysRevB.75.045133}. \\ \indent The phase
diagrams corresponding to the AF case with different value of $\phi$
are shown in Fig.\ref{polarferromag}(a). It shows that at low value of magnetic
field strength, that is $\phi=0.05$, the MI-SF phase boundary for the odd
MI lobes shifts towards larger $t/U_{0}$, indicating a stabilization of
the insulating phase while the same for the even MI lobes shows a decrease with
$t/U_{0}$. In this case, for even MI lobes, the singlet pair formation
continues to play a dominant role as pointed out in
Ref.\cite{PhysRevA.70.043628}.  \\\indent At large magnetic field
strengths, say for example, $\phi=0.1$, we found that the even MI
lobes shrink noticeably while the other MI lobes are enhanced
significantly, indicating further increase of the location for the
MI-SF phase transition at this value of the field strength. This is
quite interesting because the MI phase becomes more stable compared to
the SF phase in comparison to the results obtained in
Refs.\cite{PhysRevA.68.043612,Uesugi}, where the effect of magnetic
field enters in the Hamiltonian {\it{only}} through the zeeman interaction term. There
it was observed that the insulating phase vanishes, pushing the system
towards a SF region with increasing field strengths. In this work, at
large magnetic fields, the zeeman interaction strength pushes the
system towards the SF regime, while the magnetic flux included through the
hopping term moves the system towards the MI regime. We explain this feature more
clearly in the following discussion.
\\\indent The Bose gas in presence of a magnetic field has an analogy with
that of electrons on a thin film which suffer weak localization
effects in a disordered enviorment \cite{BERGMANN}. Weak localization
arises due to quantum interference of the paths traced by the
conduction electrons scattered off the impurities. The presence of the
magnetic field now introduces a relative phase difference arising
among the time reversed paths, that is between the two interfering waves. This
phase shift is random, and hence the magnetic field destroys the
chorence of the interfering waves, thereby suppressing the interference
pattern after a flight time proportional to $1/B$. Similar to the
conduction electrons, for Bose systems, the magnetic flux tries to
destroy the phase coherence of the SF order parameter near the
transition point causing the system to move towards the MI regime.
This explain a shift of the location for the MI-SF phase transition to
larger value of $t/U_{0}$ with increasing magnetic flux present in the
hopping term.  \\ \indent Also the instability of the even MI lobes at
higher magnetic flux values can be understood from the following
discussion. At low magnetic fields, the formation of spin
singlet (nematic) pair corresponding to the even (odd) MI lobes still
continues and hence the ground state for even MI lobes is $|0,0,n
\rangle$, while for the odd MI lobes, it is $|1,S_{z},n \rangle$. But
at high field strengths, the ground state now changes from $|0,0,n
\rangle$ to $|2,S_{z},n \rangle$ for the even MI lobes and $|1,S_{z},n
\rangle$ to $|3,S_{z},n \rangle$ for the odd MI lobes, that is
from $S$ to $S+1$ since the formation of singlet or the nematic pair no
longer occur due to the change in ground state structure at higher value of the 
magnetic field strength in the zeeman term \cite{PhysRevA.68.043612,Uesugi}.
\\ \indent The phase diagrams for the ferromagnetic case at
$U_{2}/U_{0}=0.0$ is shown in Fig.\ref{polarferromag}(b). For the
ferromagnetic interaction, the phase diagrams are similar to that of
the spin-0 (scalar) system, except for only the chemical potential width
now gets rescaled with the zeeman interaction strength, $\eta$ as
$\mu+\eta \rightarrow \mu {'}$
\cite{Uesugi,PhysRevB.75.045133}. Unlike the antiferromagnetic case,
in the ferromagnetic case, all the MI phases become more stable with
increasing magnetic field strength due to the phase decoherence of the
SF order parameter at the transition point as discussed above. Further
each MI lobe now gets shifted along the vertical axis $(\mu/U_{0})$ by
an amount $\eta/U_{0}$ due to rescalling of the chemical
potential. The phase diagram at $\phi=0.1$ are in agreement with the
results obtained in Ref.\cite{PhysRevB.75.045133}.
\begin{figure}[!ht]
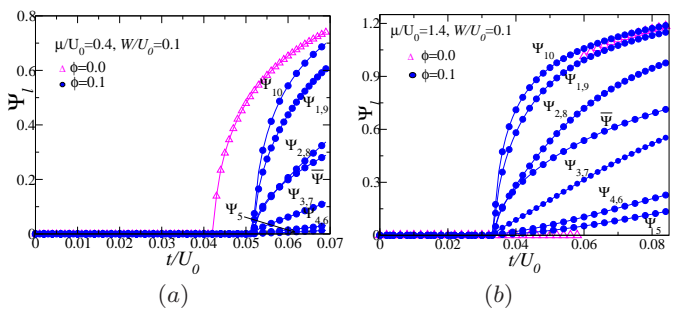

  \centerline{ \hfill
\psfig{file=2a.eps,width=0.25\textwidth}
    \hfill \hspace{0.7mm}\hfill 
      \psfig{file=2b.eps,width=0.233\textwidth}           
         \hfill} \centerline{\hfill $(a)$ \hfill\hfill $(b)$
    \hfill}
  \caption{(Color online) The 1D behaviours of $\psi_{l}$ in AF case
    corresponding to odd MI lobe (a) and even (b) are shown with
    $\phi$. The MI-SF phase transition for odd MI lobe ($\mu/U_{0}=0.4$) is second order
    for all values of $\phi$ while for even MI lobes ($\mu/U_{0}=1.4$), it is first
    order only when $\phi\le U_{2}/U_{0}$ but shows second order for higher
    magnetic flux $i.e$ $\phi > U_{2}/U_{0}$. }
\label{psil1d}
\end{figure}
\\\indent In Fig.\ref{psil1d}, we study the one dimensional behaviour
of the SF order parameter corresponding to the even and the odd MI
lobes in the antiferromagnetic case for different values of $\phi$. It
is seen that the location of the MI-SF phase transition occurs at the same value of the hopping strength, $t_{c}/U_{0}$ corresponding to different
lattice sites $l$ (of the magnetic supercell of dimension $1\times q$)
and obeys the periodicity condition, $\psi_{l}=\psi_{l+q}$.
Interestingly, for the odd MI lobes ($\mu/U_{0}=0.4$), the MI-SF phase
transition is second order in nature due to the continuous variation of the SF order parameter
and increase in $t_{c}/U_{0}$ value with increasing magnetic field
strengths [Fig.\ref{psil1d}(a)]. While for even MI lobes
($\mu/U_{0}=1.4$), for low magnetic field strengths, that is $\phi \le
U_{2}/U_{0}$, the MI-SF phase transition has a first order character
due to jump in the order parameter. However for higher field strengths,
that is $\phi > U_{2}/U_{0}$, the order parameter shows continuous variation from the
MI to the SF phase and hence shows a second order transition for all
$\psi_{l}$. Also the critical tunneling strength, $t_{c}/U_{0}$
decreases with increasing magnetic flux values because of the absence
of singlet pair formations [Fig.\ref{psil1d}(b)]. We have also studied
$\psi_{l}$ in the ferromagnetic case as a function of flux, $\phi$ and
the results are in agreement with those obtained in
Ref.\cite{PhysRevB.75.045133}.
\subsection{Three body interaction potential} 
\begin{figure}[!h]
  \centerline{ 
	\hfill \psfig{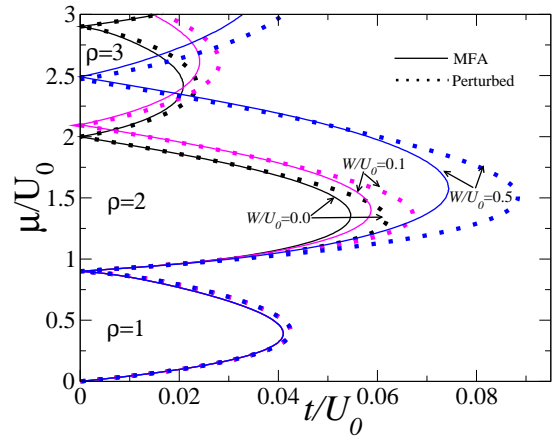} 
	\hfill
    \hfill} 
  \caption{(Color online) Phase diagrams in the AF case with
    $U_{2}/U_{0}=0.05$ for $W/U_{0}=0.1$ and $0.5$. The solid lines are for
    the mean field results and the dotted lines are obtained via
    perturbeted method.}
\label{phasepolar}
\end{figure}

\begin{figure}[!ht]
  \centerline{ 
	\hfill \psfig{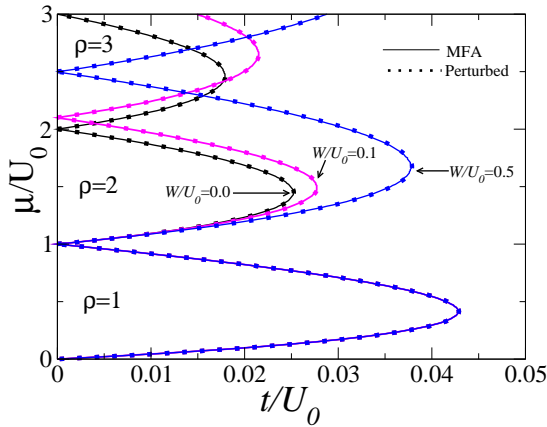} 
	\hfill
    \hfill} 
  \caption{(Color online) Phase diagrams corresponding to the
    ferromagnetic case with $U_{2}/U_{0}=0.0$ for $W/U_{0}=0.1$ and
    $0.5$. The solid lines are for the mean field results and the
    dotted lines are obtained via perturbeted method.}
\label{phaseferro}
\end{figure}

\begin{figure}[!h]
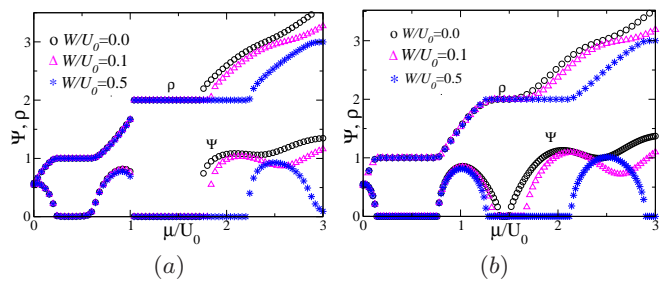

  \centerline{ \hfill
    \psfig{file=6a.eps,width=0.235\textwidth}
    \hfill \hfill \hspace{1mm}\psfig
           {file=6b.eps,width=0.235\textwidth}
           \hfill} \centerline{\hfill $(a)$ \hfill\hfill $(b)$ \hfill}
  \caption{(Color online) The variation of $\psi$ and $\rho$ with
    $W/U_{0}$ for the antiferromagnetic case is shown in (a) and the
    ferromagnetic case is shown in (b).}
\label{psirho}
\end{figure}

\indent We are also keen to explore the effect of three body
interaction on spin-1 Bose gas which enters via the Hamiltonian (in
addition to the Eq.(\ref{bhm1})) as in the following
\cite{PhysRevA.88.023602},
\begin{equation}
H_{3}=\frac{W}{6}\sum\limits_{i}n_{i}(n_{i}-1)(n_{i}-2)+\frac{V}{6}\sum\limits_{i}({\bf{S}}^{2}_{i}-2n_{i})(n_{i}-2)
\label{bhm2}
\end{equation}
where $W$ and $V$ are the three body spin independent and dependent
interaction strengths. It was found that the three body interaction
strength is related with the two body interaction strength as, $W \propto
(V_{0}/E_{r})^{3/4}a^{2}_{s}k^{2}U^{2}_{0}$ ($a_{s}$: $s$ wave scattering length and $k$: wave vector) \cite{PhysRevA.78.043603} and experimentally $a^{2}_{s}k^{2}$ is
in the order of $10^{-2}$ to $10^{-8}$ \cite{Pethick}. Thus it is reasonable to consider $W<<U_{0}$ and the relationship,
$V/U_{0}=2(U_{2}/U_{0})(W/U_{0})$ only holds for $W<<U_{0}$ and
$V<<U_{2}$ \cite{PhysRevA.88.023602}. \\ \indent With these in hand, we study the effect of {\it{only}} the three body interaction before we go on to explore the consequences of magnetic field therein
on the SF-MI phase transition. \\\indent Let us consider the atomic
limit, that is $t=0$ on the spinor BHM in Eq.(\ref{bhm2}) without a
magnetic field. At $t=0$, the Hamiltonian consists only the
unperturbed terms as,
\begin{eqnarray}
H^{0}&=&\frac{W}{6}n(n-1)(n-2)+\frac{V}{6}[S(S+1)-2n](n-2)\nonumber\\
&-&\mu n+\frac{U_{0}}{2}n(n-1)+\frac{U_{2}}{2}[S(S+1)-2n]
\end{eqnarray} 
which has a common eigenstate $|S,S_{z},n\rangle$ where the
corresponding operators, namely $S,S_{z},n$ commute with each other
and we may remove the site index $(l,m)$ for the homogeneous case. In
the atomic limit, the system is completely in the MI phase with an
energy gap, $E_{g}$ in the particle hole excitation spectra, which is
the difference between the upper ($\mu_{+}$) and lower ($\mu_{-}$)
values of the chemical potential corresponding to a MI lobe with
occupancy $n$ \cite{PhysRevB.40.546}. The $\mu_{\pm}$ can be
calculated from the following relation as
$E^{0}(S_{1},n_{1})<E^{0}(S,n)<E^{0}(S_{2},n_{2})$ where $E^{0}$ is
the eigenvalue of the $H^{0}$ and $S_{1,2},n_{1,2}$ are the lower and
higher spin and density values respectively corresponding to the $S,n$
values. Following the calculation carried out in
Ref.\cite{PhysRevA.83.013605}, this inequality corresponding to the
antiferromagnetic case leads to following conditions, which are stated
 below. \\(i) For the odd MI lobes ($n=1,3,...$):
$(n-1)+(n-1)(n-2)W/2U_{0}+(1-n)V/3U_{0}<\mu/U_{0}<n-2U_{2}/U_{0}+n(n-1)W/2U_{0}-(n-1)V/U_{0}$. If
we equate these two $\mu$ values, we shall obtain a critical
$U_{2}/U_{0}$, given by $U^{c}_{2}/U_{0}=1/2+(n-1)[W/2U_{0}-V/3U_{0}]$ below
which the odd MI lobes exist and above which the odd MI lobes vanish.
\\(ii) For even MI lobes ($n=2,4,..$): If
$U_{2}/U_{0}<U^{c}_{2}/U_{0}$ (as above), then
$(n-1)-2U_{2}/U_{0}+(n-1)(n-2)W/2U_{0}+(2-n)V/U_{0}<\mu/U_{0}<n+n(n-1)W/2U_{0}-nV/3U_{0}$.
For $U_{2}/U_{0}>U^{c}_{2}/U_{0}$,
$n-3/2-U_{2}/U_{0}+(n-2)^{2}W/2U_{0}+2(2-n)V/3U_{0}<\mu/U_{0}<n+1/2-U_{2}/U_{0}+n^{2}W/2U_{0}-2nV/3U_{0}$.
\\ Similarly for the ferromagnetic case, since there is no distinction
between the odd and even MI lobes, for all MI lobes,
$(n-1)[1+U_{2}/U_{0}+(n-2)W/2U_{0}+(n-2)V/2U_{0}]<\mu/U_{0}<n[1+2U_{2}/U_{0}+(n-1)W/2U_{0}+(n-1)V/2U_{0}]$.
\begin{figure}[!h]
  \centerline{
    \hfill \psfig{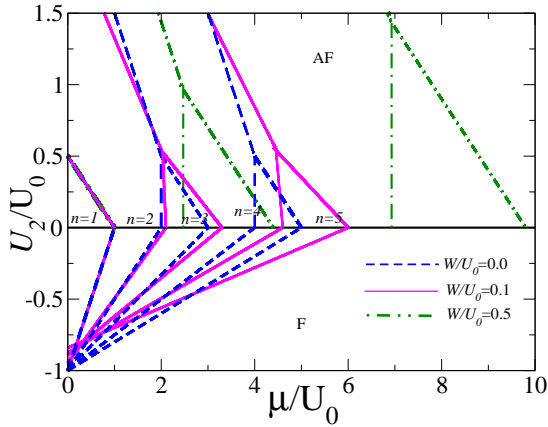}
    \hfill}
\caption{(Color online) The MI lobe structures at the atomic limit
  that is $t=0$ with $W/U_{0}$ both in the antiferromagnetic and the
  ferromagnetic cases.}
\label{miwv}
\end{figure} 
\\ \indent If we plot all these equations for different values of
$W/U_{0},V/U_{0}$, we shall obtain the MI lobe structures as shown in
Fig.\ref{miwv}. In the AF case, for $W/U_{0}=0.1$ which yields
$V/U_{0}=0.2U_{2}/U_{0}$ and we found that the even MI lobes become
more stable compared to the odd MI lobes and the width of the chemical
potential, $\mu$ for all the MI lobes, except the first one, increases
with the three body interaction strength, $W$, suggesting the dominance
of the insulating phase compared to the SF
phase. Interestingly, the critical $U^{c}_{2}/U_{0}$ for the
disappearance for all odd MI lobes in absence of $W/U_{0}$ was 0.5
\cite{PhysRevA.83.013605} now changes to 0.53 and 0.553 at
$W/U_{0}=0.1$ corresponding to the third and fifth odd MI lobes
respectively.\\ \indent We have also considered a higher value of the
three body interaction strength that is $W/U_{0}=0.5$, for which we
have chosen $V/U_{0}=0.05\sim U_{2}/U_{0}$ and found that the chemical
potential boundary gets enhanced, thereby making inroads for the MI
phase to be more stable. This is particularly true for the even MI lobes
where the critical value, $U^{c}_{2}/U_{0}$ increases accordingly with
$W,V$.  \\ \indent In the ferromagnetic case, the MI phase has similar
properties like a spin-0 (scalar) Bose gas and the increase of the right
boundary of chemical potential in Fig.\ref{miwv} results in the
increase in second and higher MI lobes width with inclusion of three
body interaction potential.  \\ \indent Now we turn on the hopping
strengths and present the phase diagrams obtained from mean field
approximation (MFA) (see Eq.(\ref{mf1})) in order to study the SF-MI
transition with three body interaction strengths. The phase diagrams
corresponding to the AF case for $U_{2}/U_{0}=0.05$ with different
values of $W/U_{0}$ are shown in Fig.\ref{phasepolar}. At
$W/U_{0}=0.1$, we found that although there is no change for the first
MI lobe, but the second and higher MI lobes get enhanced with
$W/U_{0}$ as seen from Fig.\ref{miwv}. With increasing the three body
interaction strength, $W/U_{0}$, the MI phase now encroaches more
towards the SF regime, pushing the system to an insulating phase
rather than a conducting phase. The phase diagram without $W/U_{0}$
is included for comparison which was studied earlier in
Refs.\cite{PhysRevA.70.043628,PhysRevB.77.014503}. Besides, the location
for the MI-SF phase transition is now occurring at higher values of
hopping strength, $t_{c}/U_{0}$ due to the presence of three body
interaction strength.
\\\indent Thus adding higher interaction strengths (such as a four body
term etc) in the Hamiltonian, the effect is that the system acquires higher
interaction energy which requires large hopping strengths to overcome
this potential blockade. We have also checked that the first and third
odd MI lobes vanish when the spin dependent interaction value
satisfies, $U_{2}/U_{0}=0.54\ge U^{c}_{2}/U_{0}$ at $W/U_{0}=0.1$ and $U^{c}_{2}/U_{0}=0.967$ for $W/U_{0}=0.5$. These numbers are in agreement
with the analytic calculations presented earlier for the atomic limit
corresponding to the disappearance of the respective MI lobes.  \\\indent The
variation of SF order parameter, $\psi$ and local density, $\rho$
corresponding to the AF case including $W/U_{0}$ are shown in
Fig.\ref{psirho}(a). The chemical potential, $\mu/U_{0}$ for the SF-MI
phase transition corresponding to the first odd MI lobe [MI(1)]
remains unaltered, while for the other MI lobes, it increases with the
three body interaction strength, $W/U_{0}$. Also for the even MI
phase, the MI-SF phase transition still has a first order character
and for the odd MI phase, it is a second order phase transition as
ascertained earlier in Ref.\cite{PhysRevB.77.014503} without
$W/U_{0}$.  \\ \indent The mean field phase diagrams corresponding to
the ferromagnetic case with $U_{2}/U_{0}=0.0$ is shown in
Fig.\ref{phaseferro}. It has similar phase properties like a scalar
Bose gas. As seen from Fig.\ref{miwv}, since the chemical potential
width, $\mu/U_{0}$ now increases with the three body interaction
strength for the second ($n=2$) and further higher order MI lobes,
thus all the MI lobes, except the first one, occupy more and more
space in the phase diagram compared to the SF phase. Also the critical
hopping strength, $t_{c}/U_{0}$ for the SF-MI transition increases
with $W/U_{0}$. The behaviour of $\psi$ and $\rho$ also shown in
Fig.\ref{psirho}(b) where the value of the chemical potential
increases with $W/U_{0}$ and the SF-MI phase transition still remains
second order in presence of the three body interaction
\cite{PhysRevB.77.014503}.
\\ \indent Now we shall do a perturbation calculation to provide a
strong support for all of these mean field phase diagrams computed
numerically. Further we shall ascertain the location of the MI-SF
phase transition including the three body interaction potential,
$W/U_{0}$. After applying the mean field approximation, the perturbed
Hamiltonian is given by
$H^{'}=-t\sum\limits_{\sigma}[a^{\dagger}_{\sigma}\psi_{\sigma}+h.c]+t\sum\limits_{\sigma}\psi^{2}_{\sigma}.$
Using the same eigenstate as that of $H^{0}$, the change in the ground
state energy, after incorporating the first and second order
corrections, can be expressed in a series expansion for $\psi$ as,
\begin{eqnarray}
\hspace*{-3mm}E_{g}(\psi)&=&E^{0}+E^{1}+E^{2}\nonumber\\
&\hspace*{-9mm}=&\hspace*{-5mm}E^{0}+A_{2}(U_{0},U_{2},\mu,n,W,V)\sum\limits_{\sigma}\psi^{2}_{\sigma}+O(\psi^{4})
\end{eqnarray}
\\where the coefficient, $A_{2}(U_{0},U_{2},\mu,n,W,V)$ includes the first and second
order corrections for a particular spin component $\sigma$. Minimizing
the ground state energy with respect to $\psi$ leads to
$A_{2}(U_{0},U_{2},\mu,n,W,V)=0$ and this equation yields the boundary
between the SF to MI phases.  \\\indent In the AF case, for the even
MI lobes, using a non degenerate perturbation theory as done in
Ref.\cite{PhysRevA.70.043628}, the SF-MI phase boundary can be
obtained via $A_{2}(U_{0},U_{2},\mu,n,W,V)=0$. The above equation
gives
\begin{eqnarray}
t^{-1}&=&\frac{n/3}{\mu+2U_{2}-(n-1)U_{0}-(n-2)[(n-1)W/2+V]}\nonumber \\
&+&\frac{(n+3)/3}{nU_{0}-\mu+n(n-1)W/2-nV/3}
\end{eqnarray}
\\ Similarly for the odd MI lobes, one gets,
\begin{eqnarray}
t^{-1}&=&\frac{(n+2)/3}{\mu-(n-1)[U_{0}+(n-2)W/2-V/3]}+\nonumber
\\ &&\hspace*{-8mm}\frac{4(n-1)/15}{\mu+3U_{2}-(n-1)[U_{0}+(n-2)W/2]-(4n-10)V/3}\nonumber
\\ &+&\frac{(n+1)/3}{-\mu+nU_{0}-2U_{2}+(n-1)[nW/2-V]}\nonumber
\\ &+&\frac{4(n+4)/15}{-\mu+U_{2}+nU_{0}+(n^{2}-n)[W/2-2V/3]}
\end{eqnarray}
\\ \indent If we plot these two equations with different values of $W,V$ at
$U_{2}/U_{0}=0.05$ corresponding to the MI lobe with occupancy, $n$ we
obtain the phase diagrams as shown in
Fig.\ref{phasepolar}. At $W/U_{0}=0.1$, we found that the phase
diagrams obtained using this perturbation approach are in good
agreement with the mean field phase diagrams deep inside the MI
lobes. However near the tip of the MI lobes, the mean field and phase
diagrams obtained via this technique differ from each other. This is
because the mean field approach is not a very appropriate tool to
handle fluctuations and are in fact quite inadequate at the transition
point for the MI-SF phase boundary \cite{PhysRevA.70.043628} and the
deviation increases with increasing $W/U_{0}$.  \\ \indent If we solve
the above equations, which are quadratic in $\mu$ shows that the
critical hopping strength, $t_{c}/U_{0}$ (by equating $\mu_{+}$ and
$\mu_{-}$) which denotes the location for the MI-SF phase transition is now
a function of $W$ and $V$ and increases with the three body
interaction strength, $W$.  \\\indent Similarly in the ferromagnetic
case, we have performed similar perturbation calculation and at
$U_{2}/U_{0}=0$ the obtained phase diagrams are in complete agreement
with the mean field ones corresponding to different values of
$W/U_{0}$ and they are shown in Fig.\ref{phaseferro}.
\begin{figure}[!h]
  \centerline{ \hfill
    \psfig{file=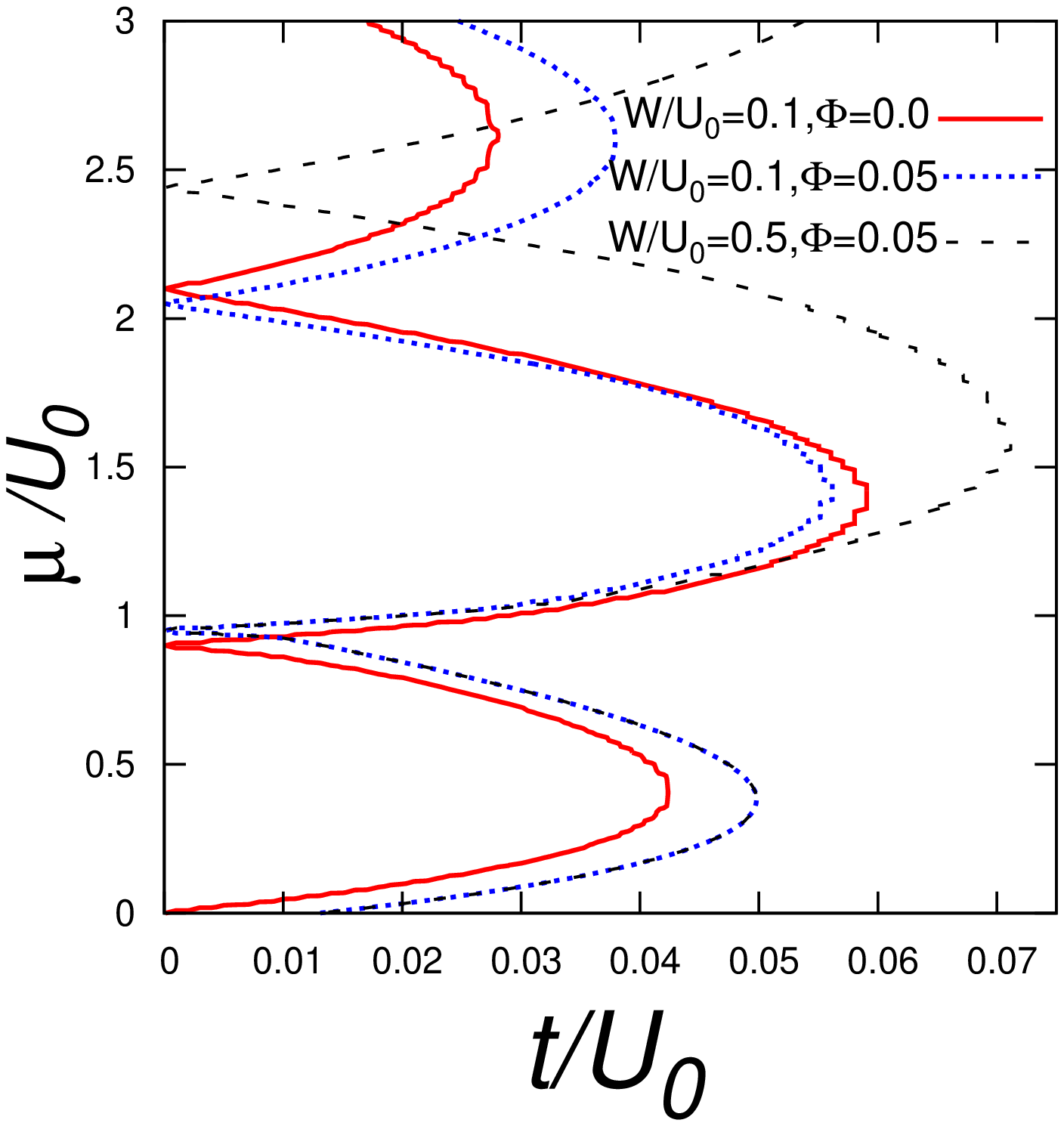,width=0.24\textwidth}
    \hfill\hspace{0.1mm} \hfill \psfig
           {file=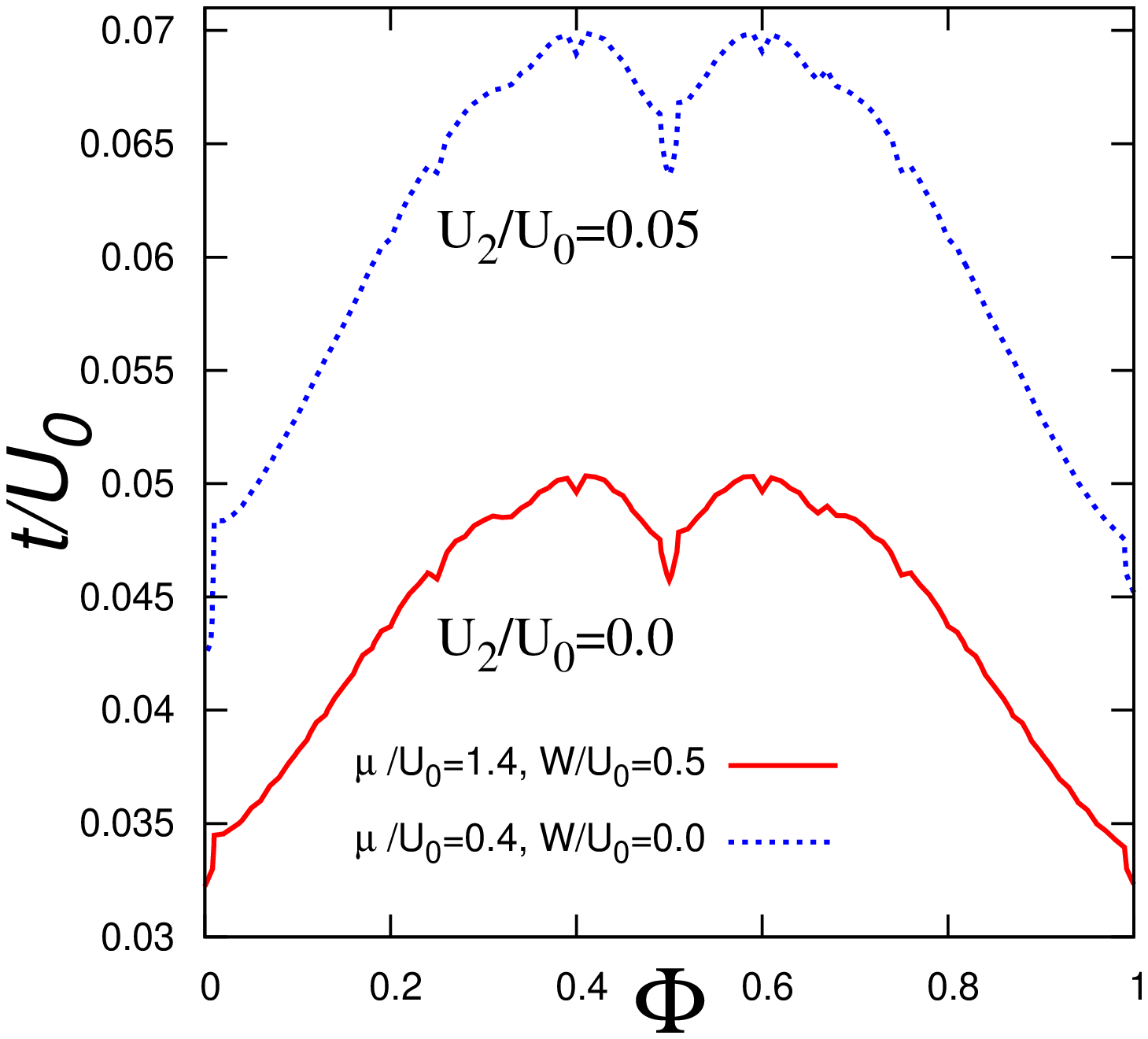,width=0.24\textwidth}
           \hfill} \centerline{\hfill $(a)$ \hfill\hfill $(b)$ \hfill}
  \caption{(Color online) Phase diagrams in the antiferromagnetic case
    with $U_{2}/U_{0}=0.05$ and $W/U_{0}$ for different values of
    $\phi$ is shown in (a). Also the phase diagram for complete range
    of $\phi$ in the antiferromagnetic and ferromagnetic cases is
    shown in (b) shows a mirror symmetry about $\phi=0.5$.}
\label{polar3flux}
\end{figure}
\\\indent Finally we incorporate the effect of the magnetic field and
compute the phase diagrams in presence of a three body interaction
potential for different values of $\phi$ and are shown in
Fig.\ref{polar3flux}(a). It shows similar effect (as without $W$) with
increasing magnetic flux strength as discussed earlier. Here the MI
phase now experiences robustness compared to the SF phase due to the presence of
$W/U_{0}$. We have also studied the SF-MI phase transition in the
ferromagnetic case corresponding to $W/U_{0}$ with different values of
$\phi$.  Another interesting property that we have obtained is
the symmetry of phase diagram as a function of $\phi$. The energy
spectrum is identical for $\phi$ and $N+\phi$ where $N$ is an
integer and is symmetric under $\phi=-\phi$ as studied in
Ref.\cite{PhysRevB.14.2239}. This is seen in Fig.\ref{polar3flux}(b),
where we consider the flux over a period of [0,1] which shows a
reflection symmetry around $\phi=0.5$ both in the antiferromagnetic
and the ferromagnetic cases and in agreement with results in
Ref.\cite{PhysRevB.75.045133} for the ferromagnetic case.
\section{Conclusion}
In this work, we have elaborately studied the effect of an external
magnetic field and a repulsive three body interaction potential on
spin-1 ultracold Bose gas. At first, we have obtained the phase
diagrams in presence of magnetic field corresponding to both types of
spin dependent interactions. In the AF case, at low magnetic field
strengths, the even MI lobes continue to play a dominant role compared
to the odd MI lobes due to formation of spin singlet pairs. While at
higher magnetic field strengths, the zeeman interaction term
suppresses the singlet pairs formation and thereby destabilizes the
even MI lobes. However the effect of magnetic field in the hopping
term, included via $e^{if(\phi)}$ ($\phi$ being the flux), pushes the
system towards the MI regime. As a result, the odd MI lobes encroach
into the SF regime. In the ferromagnetic case, the phase diagrams are
similar to that of the scalar Bose gas and the system is more likely to be in
the MI regime compared to the SF phase with increasing flux
strengths. Also the nature of MI-SF phase transition for the even MI
lobes is first order as long as flux is less than the spin dependent
interaction (scaled by $U_{0}$) but changes over to a second order transition for higher
flux strengths in the AF case. 
\\\indent In experiments, flux can be controlled by choosing the ratio of the wavenumber of the laser beams, $k_{L}$ which form the optical lattice potential to that of the Raman laser beams, $k_{R}$ which couple the internal atomic states that is $\phi=k_{R}/k_{L}$ \cite{PhysRevLett.112.043001,Stuhl1514}. Thus a desired value of the flux can be obtained by choosing the respective wavelengths accordingly. \\\indent Next we consider the effect
of a three body interaction strength without the magnetic field and
found that the chemical potential width is enhanced and hence the MI phase
occupies more region compared to the SF phase. In the AF case, the
even MI lobes become more stable compared to the odd MI lobes and the
odd MI lobes vanish when the spin dependent interaction term is
greater than a certain critical value.
The location of the critical tunneling strength for the MI-SF phase
transition also increases (towards larger $t/U_{0}$) with the strength
of the three body interaction term.  
\\\indent Experimentally the three and higher body
interaction terms were successfully observed using atom
interferometry as studied in Ref.\cite{Will} and photon assisted
tunneling in Ref.\cite{PhysRevLett.107.095301}. Also it has been proposed to observe three body interaction effects using an
optical lattice and superlattice potential in 
Ref.\cite{PhysRevA.85.051604}. Recently Paul $et.$ $al.$ successfully engineered a Bose Hubbard Hamiltonian with an attractive three body interaction potential which is dominant than the two body interaction potential \cite{PhysRevA.93.043616}.
\\\indent A perturbation
calculation has also been done to provide a support for the mean field
phase diagrams. In the AF case, the phase diagrams obtained via
perturbation calculation are in good agreement with the mean field
approach deep inside the MI regime, but they differ near the tip of
the MI lobes and the discrepancy becomes noticeable with increasing
value of three body interaction strength. While in the ferromagnetic
case, the mean field phase diagrams are in complete agreement with
those obtained using the perturbed calculation.  \\\indent We have
also studied the SF-MI phase transition with a three body interaction
term corresponding to different values of the magnetic flux and they
show similar properties as that corresponding to the case without a
three body interaction term. Besides, the system shows a reflection
symmetry about $\phi=0.5$ both for the antiferromagnetic and the
ferromagnetic cases over a period of $\phi \in [0,1]$.\\ \indent However all the experimental results cited above are relevant to
the scalar Bose gas. We have a strong conviction that our theoretical
results on the spinor Bose gas will be useful to ascertain many of the
interesting phenomena that are otherwise absent in the scalar Bose
gas.
\section*{Acknowledgments}
SNN likes to thank Prof. M. Oktel and Dr. O. Umucalilar for their
help. We thank Prof. R. V. Pai for useful discussions. SB thanks
CSIR, India for financial support under the grants F no:
03(1213)/12/EMR-II.  \bibliography{referance} \bibliographystyle{aip}
\end{document}